\documentclass[epj]{svjour}
\usepackage{graphics}
\begin{document}
\title{Backward asymmetry measurements in the elastic pion-proton scattering at resonance energies}
\author{I.G.~Alekseev\inst{1}, N.A.~Bazhanov\inst{2}, 
Yu.A.~Beloglazov\inst{3},P.E.~Budkovsky\inst{1},
E.I.~Bunyatova\inst{2}, E.A.~Filimonov\inst{3},
V.P.~Kanavets\inst{1}, A.I.~Kovalev\inst{3},
L.I.~Koroleva\inst{1}, 
B.V.~Morozov\inst{1},
V.M.~Nesterov\inst{1}, D.V.~Novinsky\inst{3}, V.V.~Ryltsov\inst{1},
V.A.~Shchedrov\inst{3}, A.D.~Sulimov\inst{1}, V.V.~Sumachev\inst{3},
D.N.~Svirida\inst{1}, V.Yu.~Trautman\inst{3} \and L.S.~Zolin\inst{2}
}                     
\institute{Institute for Theoretical and Experimental Physics, Moscow, 117218, Russia \and
Joint Institute for Nuclear Research, Dubna, Moscow area, 141980, Russia \and
Petersburg Nuclear Physics Institute, Gatchina, Leningrad district, 188300, Russia}
\date{Received: date / Revised version: date}
%
\abstract{
      The asymmetry parameter $P$ was measured for
the elastic pion-proton scattering in the very backward angular region 
of $\theta_{cm}\approx 150-170^o$ at several pion beam 
energies in the invariant mass range containing most of the
pion-proton resonances. 
    The general goal of the experimental program was to provide
new data for partial wave analyses in order to resolve 
their uncertainties 
in the baryon resonance region to allow the unambiguous baryon
spectrum reconstructions.
Until recently the parameter $P$ was not measured in the examined 
domain that might be explained by the extremely low cross section.
At the same time the predictions of various partial wave analyses
are far from agreement in some kinematic areas and specifically
those areas were chosen for the measurements where the disagreement
is most pronouncing.
    The experiment was performed at the ITEP U-10 proton synchrotron, 
Moscow, by the ITEP-PNPI collaboration in the latest 5 years.
\PACS{
      {13.88.+e}{Polarization in interactions and scattering}   \and
      {13.75.Gx}{Pion-Baryon interactions}
     } 
} 
\authorrunning{I.G.~Alekseev et al.}
\titlerunning{Backward asymmetry in elastic $\pi p$ scattering at resonance energies}
\maketitle
\section{Introduction}
\label{intro}
    The history of the investigations of the pion-nucleon interaction 
is about 40 years old. The systematic set of data in the resonance
region was obtained during this period. These days three partial 
wave analyses (PWA) are known which are the main source of the 
information on the baryon resonance parameters: the basic PWA 
of the Karlsruhe-Helsinki group KH80 \cite{c:KH80,c:KH80a} 
in the beam momentum  range 0.020$-$10~GeV/c, analysis of the 
Carnegie-Mellon-Lawrence-Berkeley-Laboratory group CMB80 \cite{c:CMB} up 
to 2.5~GeV/c and the series of solutions by George Washington
University (GWU) group, formerly located at VPI, below 2.1~GeV/c 
\cite{c:VPI,c:GWU}. 
The detailed presentation of all mentioned PWA results can be found 
at CNS DAC web page through the SAID program \cite{c:SAID}.

\begin{figure}
\begin{center}
\resizebox{0.35\textwidth}{!}{\includegraphics{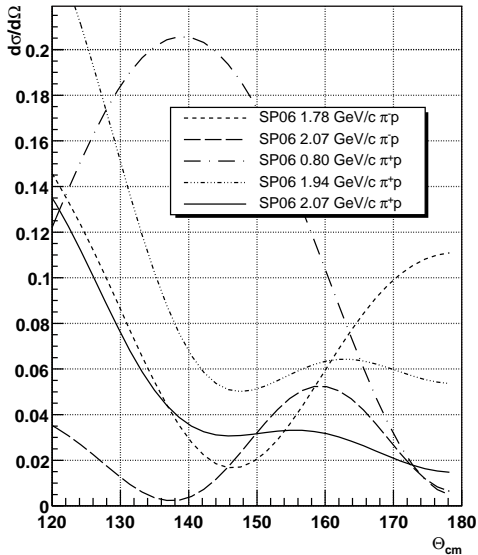}}
\end{center}
\caption{Cross section in the elastic $\pi p$ scattering in the backward
region at several beam momenta. The data are taken from SP06 PWA~\cite{c:GWU}.}
\label{fig:DCS}
\end{figure}

    But even up to now there are some kinematic regions where
the predictions of the different PWA are in evident contradiction with one 
another. In most cases this is due to the very low cross section
values of the order of 0.1~mb/sr (fig.~\ref{fig:DCS}).
Low quality or absence of experimental data in such regions
leads to the continuous uncertainties in partial amplitudes
as well as to the unresolved discrete ambiguities and consequent
wrong choice of solution branches~\cite{c:INFL}. 
These minima in the differential cross section are caused by
the specific behavior of the pion-nucleon amplitude or, rather,
of the Barrelet zero trajectories, and that is why the measurements
in such areas are particularly important for the correct
reconstruction of the pion-nucleon amplitude.
The main goal of this work is to provide new experimental information
for the partial wave analyses specifically is such kinematic
areas, where it may best help to resolve the existing PWA uncertainties
and thus allow the reliable extraction of non-strange baryon resonance 
spectrum and properties.

    This paper presents the latest results of the ITEP-PNPI collaboration
on the measurements of the asymmetry $P$ both for $\pi^+p$ and $\pi^-p$
elastic scattering to the very backward angles in several energy points.
 
\section{Experimental conditions}
 
    Polarization parameter $P$ in the elastic scattering is determined
by the direct measurement of the azimuthal asymmetry of the reaction produced
by incident pions on a proton target polarized normally to the scattering plane.
 
The differential cross section in this case has the following form:

\begin{displaymath}
d\sigma/d\Omega=({d\sigma/d\Omega)}_{0}[1+P\cdot(\vec{{P}_{T}}\cdot\vec{n})]
\end{displaymath}
\noindent
where ${(d\sigma/d\Omega)}_{0}$ is the differential cross section of 
${\pi}{p}$ elastic scattering on the unpolarized target, $P$ is 
polarization parameter, $\vec{{P}_{T}}$ is the vector of the target 
polarization and $\vec{n}$ is the normal to the scattering plane.

\begin{figure*}
\begin{center}
\resizebox{0.7\textwidth}{!}{\includegraphics{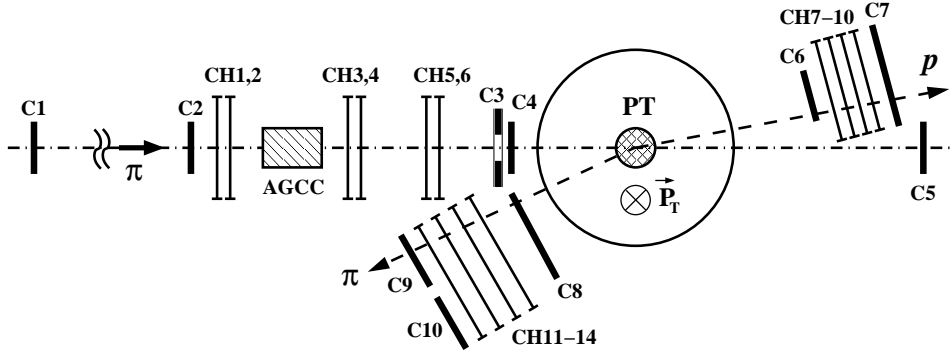}}
\end{center}
\caption{Schematic top view of the SPIN-P02 setup.}
\label{fig:SETUP}
\end{figure*}

    The experimental setup SPIN-P02 is schematically shown in 
fig.~\ref{fig:SETUP}. The 
polarized target with vertical orientation of the polarization vector
$\vec{{P}_{T}}$ is located at the focus of the secondary pion beam line
of the ITEP proton synchrotron. Scattered pion and recoil proton
are tracked with blocks of multi-wire chambers and the full event
reconstruction is performed allowing good selection of the elastic events.
The asymmetry is determined by measuring the normalized  event counts for two
opposite directions of the target polarization vector.
Such configuration corresponds to the direct measurement of the polarization 
parameter $P$ by means of the scattering asymmetry rather than determination
of the recoil proton polarization vector  with the help of a proton polarimeter. 

The main parts of the SPIN-P02 setup~\cite{c:SETUP} are:
\begin{itemize}
\item Polarized proton target PT located in the cryostat with superconducting
magnet. The container with polarized target material
(propanediole ${C}_{3}{H}_{8}{O}_{2}$ doped by ${Cr}^{V}$ complexes) 
is placed into
the magnetic field of 2.35~T created by a Helmholtz pair of 
superconducting coils. The relative free proton density is close to 
10\%~\cite{c:TARG}. 
The container has a cylindrical form with both height and 
diameter of 30~mm. Cooling of the target material to 0.5~K is provided
by an evaporation-type $^{3}He$ cryostat. The protons of the target are 
polarized by the dynamic nuclear orientation method. 
The polarization is 70-80\%   with the
measurement uncertainty about 1.5\%. The polarization sign was reversed once
per day.
\item Three blocks of two-coordinate multi-wire chambers 
with the spatial accuracy about 0.5~mm
for the tracking of the incident pions CH1-CH6, recoil protons CH7-CH10
and scattered pions CH11-CH14. In some configurations of the setup
another block was added adjacent to CH11-CH14 to achieve a wider pion
acceptance.
\item Scintillation counters C1-C12 for the trigger logic. The readout
of the chamber data was initiated on the following condition:
$C1\cdot C2\cdot \overline{C3}\cdot C4\cdot\overline{C5}\cdot 
(C6\cdot C7)\cdot (C8\cdot (C9+C10))$
\item In case of positive beam TOF technique was used to separate
pions from protons at low beam energies, while at beam momenta
above 1.8~GeV/c the aerogel Cherenkov counter AGCC~\cite{c:AEROGEL} 
was additionally introduced into the beam for the pion tagging.
\end{itemize}

    The secondary pion beam is formed by the universal magneto-optical
channel based on the two-stage achromatic scheme and provides the
intensity of up to $2\cdot10^5$ pions per accelerator spill 
$(t\simeq 0.6$~s every 4~s). The 46.5~m
long beam line is equipped with two vertical correctors allowing
to change the beam position on the polarized target with the accuracy
better than 4~mm. Beam dimensions in the target plane are 32~mm x 27~mm (FWHM). 
The momentum spread of the beam is about $\pm$2\%.

    To achieve the acceptance coverage of the very backward pion
angles about 170$^o$ at beam momentum 0.8~GeV/c two configurations
of the setup were used with opposite direction of the target
magnetic field and correspondingly different positions of the proton arm.

    The CAMAC based readout system allows to record up to 40 events 
per one accelerator cycle. Data storage as well as its on-line processing
and the setup monitoring is provided by a single PC. 

\section{Data processing}

    The processing of the data was performed in the following steps.
The straight trajectories of the beam pion, scattered pion and recoil 
proton were reconstructed in the corresponding chamber blocks outside
the magnetic field of the target by the least square method. 
Obtained tracks for each event were then extrapolated to the target region
through the magnetic field and the interaction vertex was found.
Polar and azimuthal angles were separately determined 
for both scattered particles and their values were used for the
event selection.

\begin{figure}
\begin{center}
\resizebox{0.42\textwidth}{!}{\includegraphics{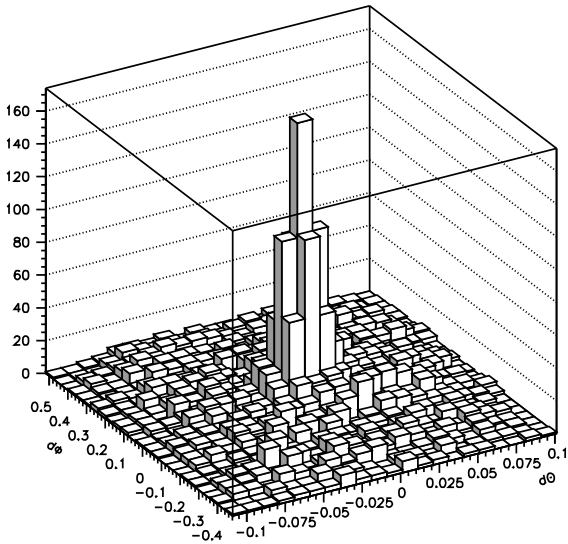}}
\end{center}
\caption{Deviations from the elastic kinematics in $\pi^+p$ scattering at 0.80~GeV/c}
\label{fig:DTDFI}
\end{figure}

    The procedure of the elastic event selection is illustrated by 
fig.~\ref{fig:DTDFI}. For each event the deviation from the elastic kinematics
was calculated in terms of two variables: $\Delta\theta$ -- the difference
in c.m. scattering angle for the pion and the proton and $\Delta\phi$ --
sum of their azimuthal deviations from the scattering plane, and 
two-dimensional distributions in these variables were filled. 
For the best background estimate the distributions were fit with a 2-dim
12-parameter polynomial excluding the area of the elastic peak and
the error matrix was calculated for the obtained fit parameters. The number
of the elastic events was determined as the distribution excess over
the background, interpolated to the area under the peak.  
To account for the different statistics with opposite target
polarization signes the intensity normalization was done based
on the quasi-elastic event counts which are believed to be unpolarized
and represent the main content of the background. Comparison
of the results with various cuts around the elastic peak allowed to
make additional systematic error tests. Statistical error accounts
for the elastic event number, the background error matrix  and the
intensity normalization uncertainty. 

    The selected  events were divided into several angular intervals 
in $\theta_{cm}$ and average angle was calculated
for each interval according to the individual values from each event.
 
\section{Results}

\begin{table}
\caption{Summary of the obtained statistics and experimental conditions}

\label{tab:SUM}
\begin{tabular}{lp{1.2cm}p{1.6cm}ll}
\hline\noalign{\smallskip}
 & $p_{\rm BEAM}$, GeV/c & $d\sigma/d\Omega$, mb/sr & 
    $N_{\rm EL}$ & $N_{\rm BG}/N_{\rm EL}$  \\ 
\noalign{\smallskip}\hline\noalign{\smallskip}
$\pi^-p$ & 1.78 & 0.02-0.10 &  3252 & 0.45  \\
$\pi^-p$ & 2.07 & 0.03-0.05 &  2047 & 0.43  \\
$\pi^+p$ & 0.80 & 0.03-0.17 &  3128 & 0.19  \\
$\pi^+p$ & 1.94 & 0.05-0.07 &  1066 & 1.01  \\
$\pi^+p$ & 2.07 & 0.02-0.03 &  701  & 1.56  \\
\noalign{\smallskip}\hline
\end{tabular}
\end{table}

    The asymmetry $P$ was measured in the $\pi p$ elastic scattering 
in the very backward angular region of $\approx(150-170)^o$ in the
c.m. frame at several beam momenta. The values of the momentum
were intensionally chosen so that the disagreement in the PWA predictions
is most pronouncing in the backward hemisphere at these momenta.
It is worth mentioning that the differential cross section
in the regions of the experiment is very different though very
small in all of them. Thus the experimental conditions in terms
of the event rate and signal to background ratio are also significantly
different. Table~\ref{tab:SUM} summarizes the statistical material
obtained for all data regions. The third column together with
fig.~\ref{fig:DCS} presents the cross section values in the angular domain
under study, giving the explanation to the number of elastic events
obtained in approximately equal running periods (column 4) and background
levels (column 5).

\begin{figure*}
\begin{center}
\begin{tabular}{cc}
\resizebox{0.31\textwidth}{!}{\includegraphics{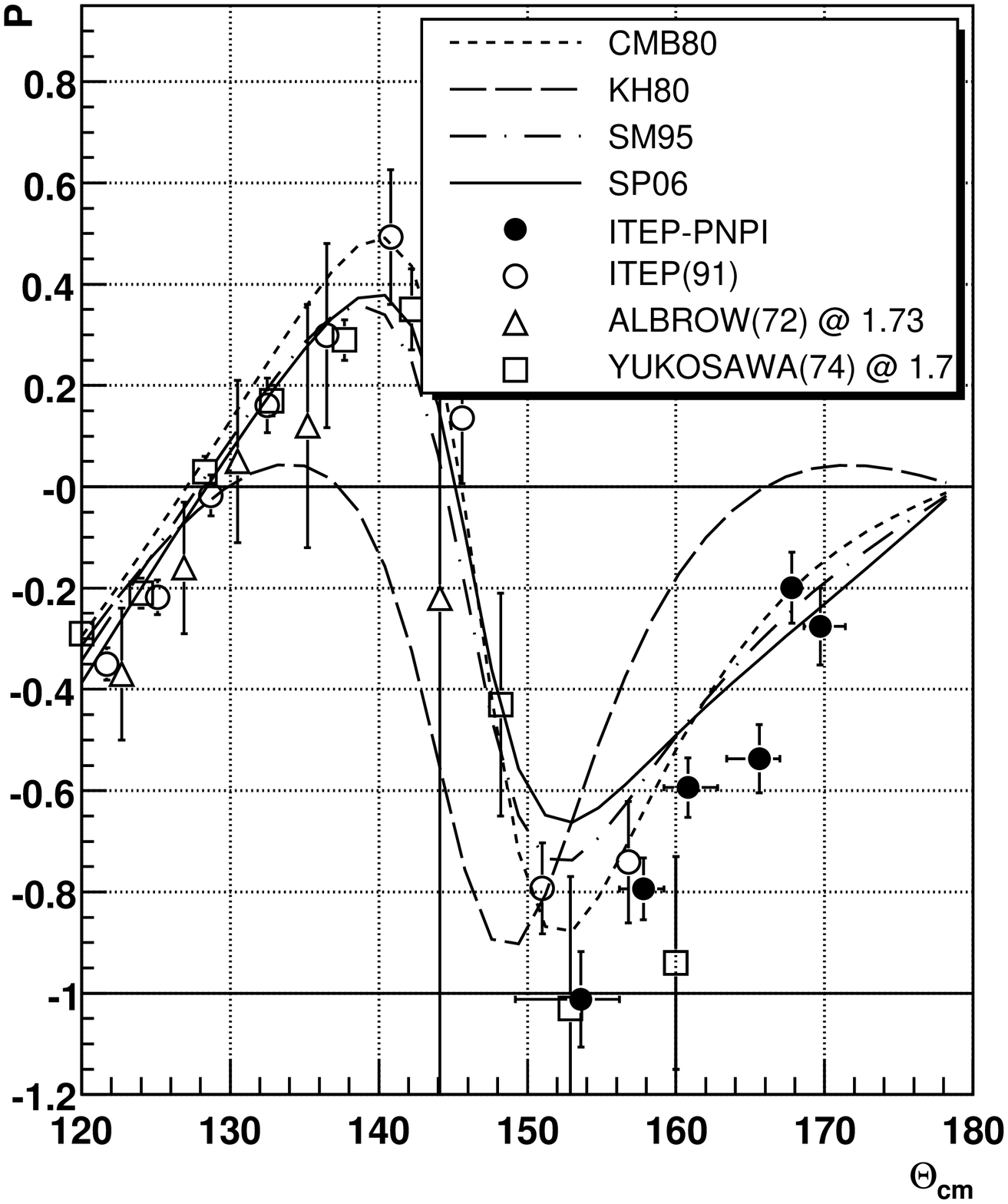}} &
\resizebox{0.31\textwidth}{!}{\includegraphics{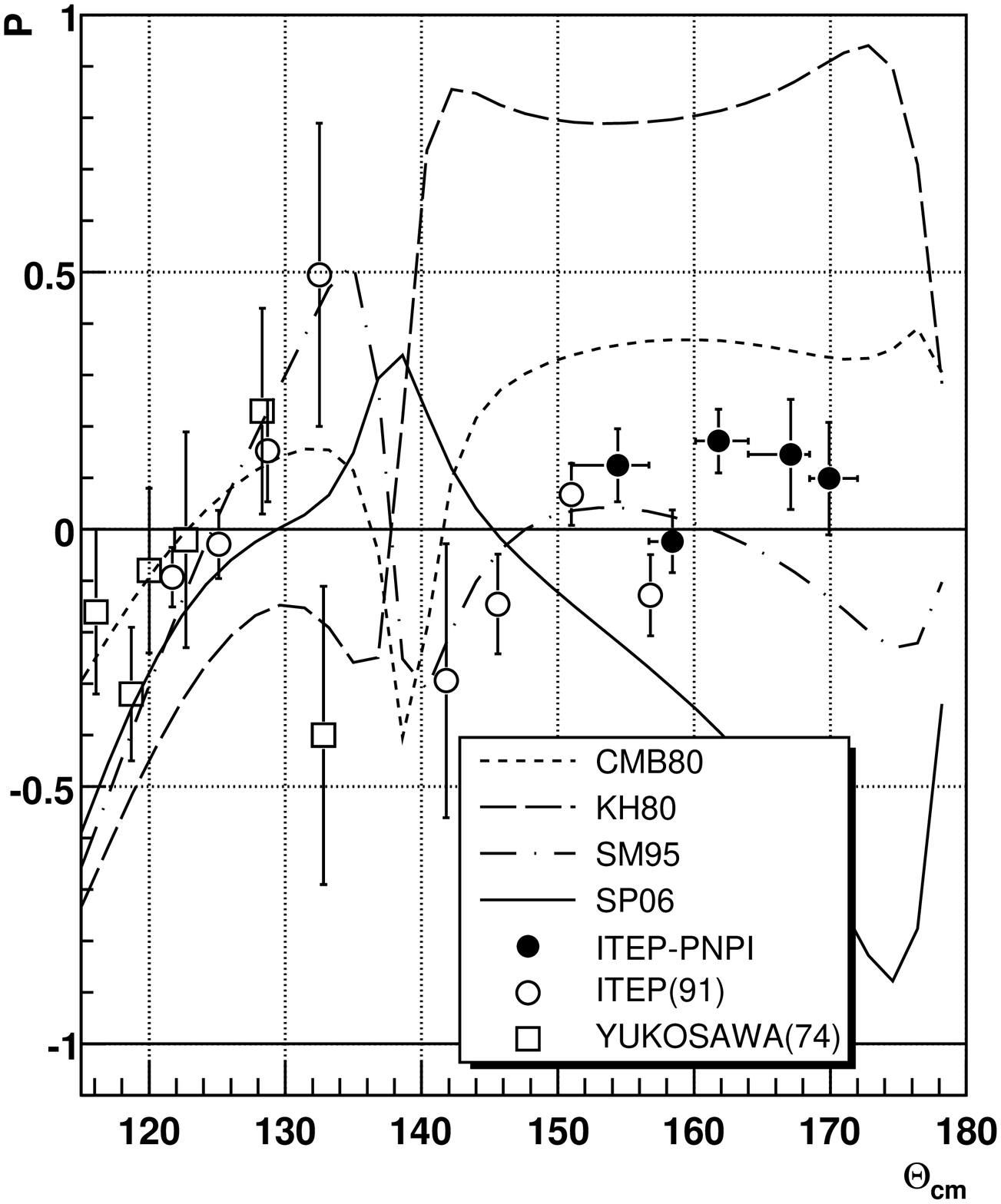}} \\
a) & b)\\
\end{tabular}
\end{center}
\caption{Asymmetry in $\pi^-p$ elastic scattering at 1.78~GeV/c (a) 
and 2.07~GeV/c (b). Earlier data are from \cite{c:ITEP} (open circles),
\cite{c:ALBROWM} (triangles) and \cite{c:YUKOSAWA} (squares).}
\label{fig:PMINUS}
\end{figure*}

\begin{figure*}
\begin{center}
\noindent
\begin{tabular}{ccc}
\resizebox{0.31\textwidth}{!}{\includegraphics{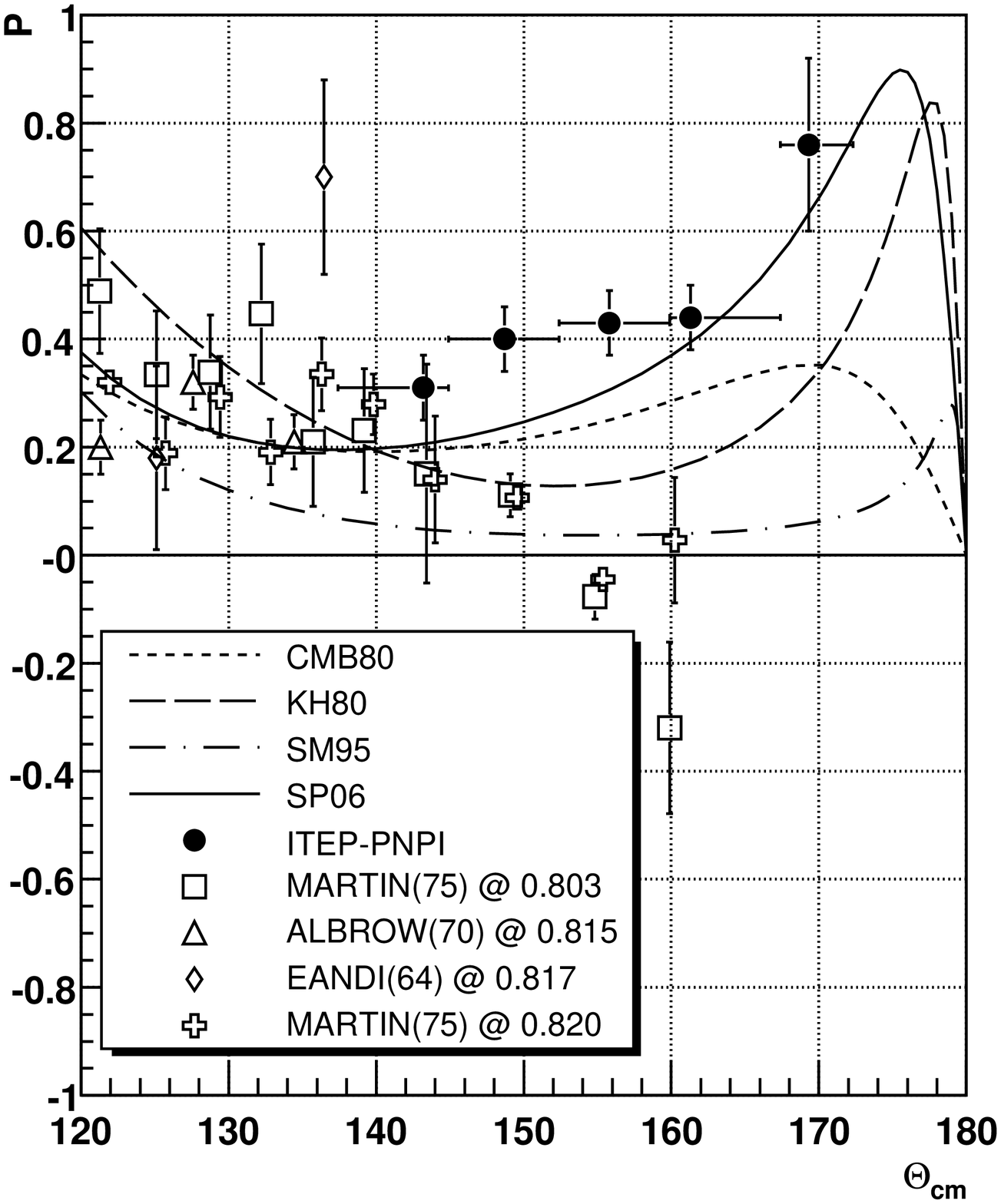}} &
\resizebox{0.31\textwidth}{!}{\includegraphics{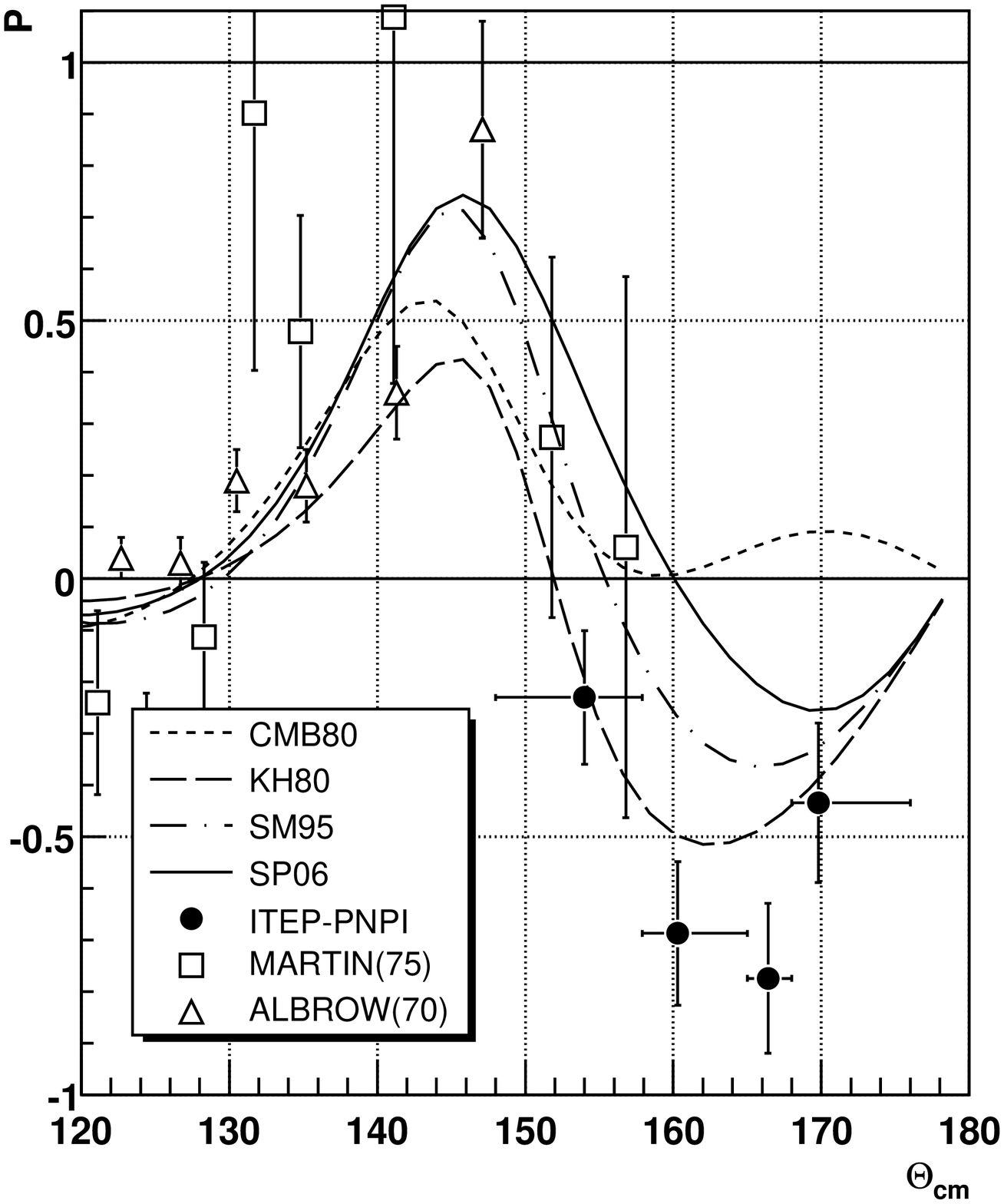}} &
\resizebox{0.31\textwidth}{!}{\includegraphics{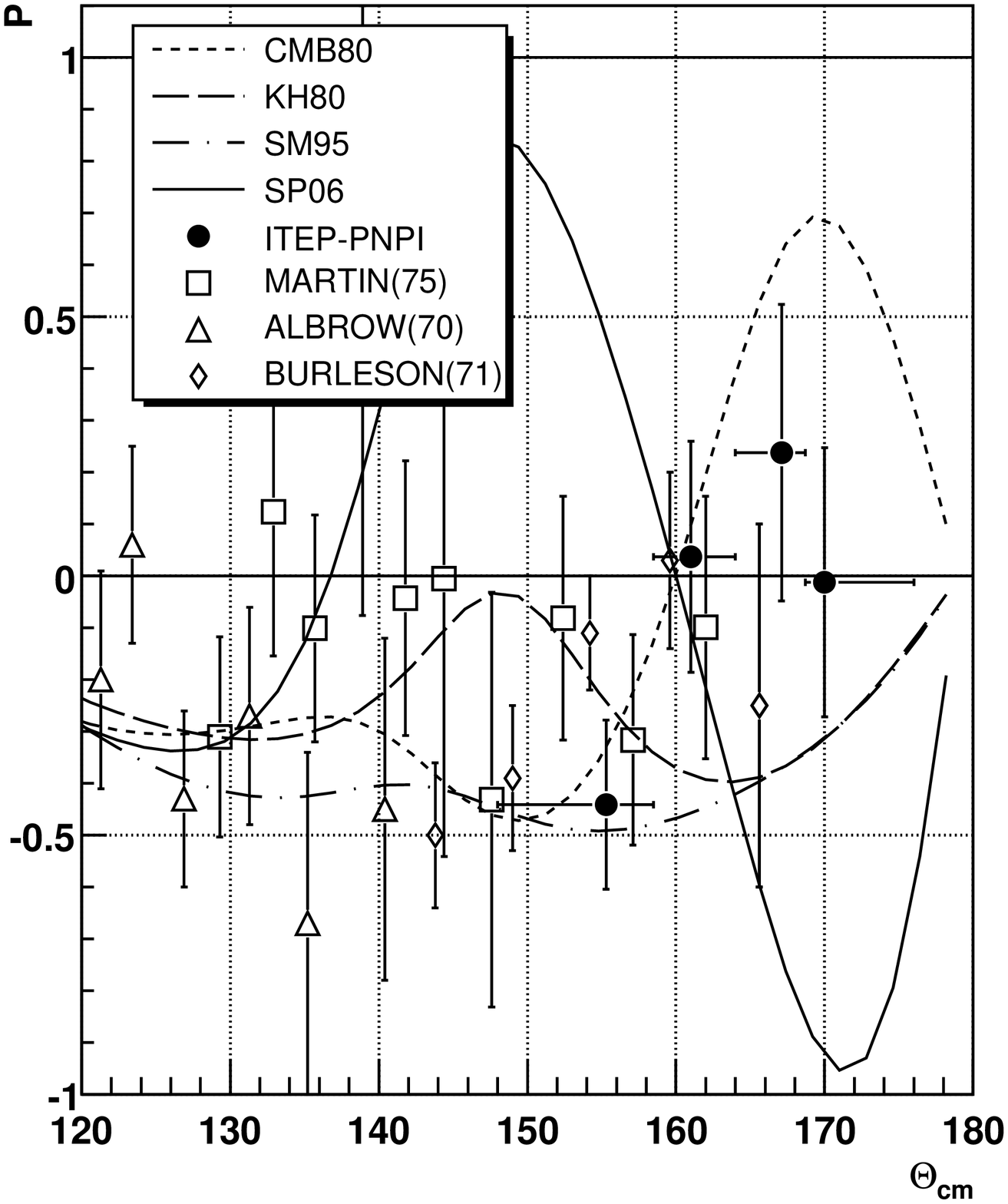}} \\
a) & b) & c) \\
\end{tabular}
\end{center}
\caption{Asymmetry in $\pi^+p$ elastic scattering at 0.80~GeV/c (a), 
1.94~GeV/c (b) and 2.07~GeV/c (c). Earlier data are from \cite{c:MARTIN}
(squares and crosses), \cite{c:ALBROWP} (triangles) and \cite{c:EANDI},
\cite{c:BURLESON} (diamonds).}
\label{fig:PPLUS}
\end{figure*}

The tables~\ref{tab:P178M}-\ref{tab:P207P} together with 
figs.~\ref{fig:PMINUS},\ref{fig:PPLUS} present the results of the measurements
as a function of the c.m. scattering angle. The errors are only 
statistical while the systematic scale uncertainty due to the target
polarization measurement is below 3\% (typically 1.5\%).
The lines in the figures show the PWA predictions for comparison:
"classic" analyses CMB80~\cite{c:CMB} (dotted) and KH80~\cite{c:KH80} 
(dashed), the latest solution of GWU group SP06~\cite{c:GWU} (solid)
and their earlier solution SM95~\cite{c:VPI} (dash-dotted).
Older experimental data at smaller angles and in the overlapping 
regions are also shown with open markers.

\begin{table}
\caption{Asymmetry in $\pi^-p$ elastic scattering at 1.78~GeV/c}
\label{tab:P178M}
\begin{tabular}{llll}
\hline\noalign{\smallskip}
Mean angle & Interval of angles & Asymmetry & Error  \\ 
\noalign{\smallskip}\hline\noalign{\smallskip}
153.6 & 149.2-156.2 & -1.012 & 0.094  \\
157.8 & 156.2-159.2 & -0.794 & 0.061  \\
160.8 & 159.2-162.8 & -0.594 & 0.059  \\
165.6 & 163.4-167.0 & -0.537 & 0.067  \\
167.8 & 167.0-168.6 & -0.199 & 0.070  \\
169.7 & 168.6-171.4 & -0.276 & 0.076  \\
\noalign{\smallskip}\hline
\end{tabular}
\end{table}

\begin{table}
\caption{Asymmetry in $\pi^-p$ elastic scattering at 2.07~GeV/c}
\label{tab:P207M}
\begin{tabular}{llll}
\hline\noalign{\smallskip}
Mean angle & Interval of angles & Asymmetry & Error  \\ 
\noalign{\smallskip}\hline\noalign{\smallskip}
154.4 & 151.0-156.7 &  0.125 & 0.071  \\
158.4 & 156.7-158.4 & -0.023 & 0.061  \\
161.8 & 160.1-164.0 &  0.172 & 0.062  \\
167.1 & 164.0-168.5 &  0.146 & 0.107  \\
169.9 & 168.5-172.0 &  0.099 & 0.109  \\
\noalign{\smallskip}\hline
\end{tabular}
\end{table}

\begin{table}
\caption{Asymmetry in $\pi^+p$ elastic scattering at 0.80~GeV/c}
\label{tab:P080P}
\begin{tabular}{llll}
\hline\noalign{\smallskip}
Mean angle & Interval of angles & Asymmetry & Error  \\ 
\noalign{\smallskip}\hline\noalign{\smallskip}
   143.2 &  137.4-144.9 &   0.310 & 0.060 \\
   148.7 &  144.9-152.4 &   0.400 & 0.060 \\
   155.8 &  152.4-159.9 &   0.430 & 0.060 \\
   161.3 &  159.9-167.4 &   0.440 & 0.060 \\
   169.3 &  167.4-172.3 &   0.760 & 0.160 \\
\noalign{\smallskip}\hline
\end{tabular}
\end{table}

\begin{table}
\caption{Asymmetry in $\pi^+p$ elastic scattering at 1.94~GeV/c}
\label{tab:P194P}
\begin{tabular}{llll}
\hline\noalign{\smallskip}
Mean angle & Interval of angles & Asymmetry & Error  \\ 
\noalign{\smallskip}\hline\noalign{\smallskip}
   154.0 &  148.0-157.9 &  -0.230 & 0.129 \\
   160.3 &  157.9-165.0 &  -0.687 & 0.139 \\
   166.4 &  165.0-168.0 &  -0.774 & 0.145 \\
   169.8 &  168.0-176.0 &  -0.434 & 0.154 \\
\noalign{\smallskip}\hline
\end{tabular}
\end{table}

\begin{table}
\caption{Asymmetry in $\pi^+p$ elastic scattering at 2.07~GeV/c}
\label{tab:P207P}
\begin{tabular}{llll}
\hline\noalign{\smallskip}
Mean angle & Interval of angles & Asymmetry & Error  \\ 
\noalign{\smallskip}\hline\noalign{\smallskip}
   155.3 &  148.0-158.5 &  -0.441 &  0.163 \\ 
   161.0 &  158.5-164.0 &   0.037 &  0.223 \\
   167.1 &  164.0-168.7 &   0.238 &  0.286 \\
   170.0 &  168.7-176.0 &  -0.012 &  0.260 \\
\noalign{\smallskip}\hline
\end{tabular}
\end{table}

The points at 1.78~GeV/c in $\pi^- p$ (fig.~\ref{fig:PMINUS}a) were taken 
to check the setup. 
Partial wave analyses do not show large variations from
each other. The data show best agreement with CMB80, while the
deviation of KH80 from the data points is obviously statistically
significant. The point at 153.6$^o$ coincides with earlier data
from~\cite{c:YUKOSAWA} and supports even deeper and saturated minimum
than any of the analyses show. Such minima indicates that a branching
point of a PWA solution may be present close to this kinematic point.
Next point at 157.8$^o$ overlaps with the previous ITEP measurement~\cite{c:ITEP}
with different setup, proofing the good quality of the result.

The data at 2.07~GeV/c with negative pions (fig.~\ref{fig:PMINUS}b)
are not exactly followed
by any of the PWA solutions. Yet the closest two are SM95 and CMB80,
while SP06 and KH80 do not even resemble the data behavior. Again
good agreement with earlier ITEP data in the overlapping region
should be stated. A narrow local minimum at 157$^o$ confirmed by
two independent measurements draws additional attention and
indicates the significant presence of partial wave with high
orbital momentum of the order of $L=8-9$. Similarly the sharp
step at 167$^o$ in 1.78~GeV/c data may point out to the large value
of the same partial amplitude.

In both cases with $\pi^- p$ scattering the latest solution SP06
of the GWU group is not in the closest to the new data.
On the contrary, in the lower energy domain and the pure $I=3/2$
isospin state the data are best described by this very solution
(see fig.~\ref{fig:PPLUS}a for $\pi^+ p$ at 0.80~GeV/c). CMB80
does not show the sharp and high peak at 175$^o$ implied by the data,
while SM95 gives much smaller values of the asymmetry in the whole
angular range of the measurement. It is worth mentioning that our
data are in obvious contradiction to the 3 rightmost points from~\cite{c:MARTIN}
at two adjacent energies. The authors claim very small errors
for 4 out of these 6 points while the GWU group already excluded some
of them from their analysis database. 

$\pi^+ p$ asymmetry at 1.94~GeV/c shows large negative values around 
165$^o$ (fig.~\ref{fig:PPLUS}b).
Neither of the solutions manifest so deep a minimum though all but CMB80
have qualitatively similar behavior. The closest prediction in this case
is from KH80.

The cross section for backward angles at 2.07~GeV/c is extremely low
for positive pions. That is why the quality of all data points in this
region is extremely poor and hardly allows to trace any particular
behavior of the asymmetry angular dependence. Newly obtained
results are not much better statistically and feature the background
levels higher than the useful elastic event numbers (table~\ref{tab:SUM}).
Yet they allow to make some conclusions about their correspondence
to various PWA solutions. The angular dependence of the data most resemble
the curve from CMB80 analysis. All other solutions show qualitatively
different behavior though KH80 and SM95 are not beyond $3\sigma$
boundary of the data.

\section{Conclusions}

The obtained results show that in some kinematic areas one or both
of the "classic" partial wave analyses CMB80 and KH80 are in
disagreement with the new data. In some cases even the qualitative
behavior of mentioned PWA does not correspond to that of the data,
which may indicate the wrong choice of the solution branch by
these analyses and, consequently, wrong extraction of baryon properties.
The latest solution SP06 of GWU group seems to be consistent with
the data in the lower energy domain, while in the region around
2~GeV/c beam momentum it's behavior looks unstable.

The ITEP-PNPI experimental team believes that their new data
on the backward asymmetry in the elastic pion-proton scattering
notably improves the database for partial wave analyses and helps
to make another step on the way of the elimination of PWA ambiguities
of various kinds and thus obtain reliable light baryon spectrum.

\section{Acknowledgments}
Our thanks to I.~Strakovsky for the interest in
our experiment and fruitful discussion on the subject. 
We are grateful to the ITEP accelerator staff for providing us with
the beam of excellent quality.

The work was partially supported by Russian Fund for Basic Research
(grants 02-02-16121-a and 04-02-16335-a), Russian State Corporation
on the Atomic Energy 'Rosatom'
and Russian State program "Fundamental Nuclear Physics".


\begin{thebibliography}{12}
\bibitem{c:KH80}
G.~H\"oehler, \textit{Handbook of Pion Nucleon Scattering, Physics Data
No. 12-1} (Fachinformationzentrum, Karlsruhe 1979).
\bibitem{c:KH80a}
G.~H\"oehler \textit{et al.}, $\pi N$-Newsletter \textbf{9}, 1 (1993).
\bibitem{c:CMB}
R.E.~Cutcosky \textit{et al.}, Phys. Rev. D \textbf{20}, 2839 (1979).
\bibitem{c:VPI}
R.A.~Arndt \textit{et al.}, Phys. Rev. C \textbf{52}, 2120 (1995).
\bibitem{c:GWU}
R.A.~Arndt \textit{et al.}, Phys. Rev. C \textbf{69}, 035213 (2004); \\
R.A.~Arndt \textit{et al.}, Phys. Rev. C \textbf{74}, 045205 (2006).
\bibitem{c:SAID}
R.A.~Arndt \textit{et al.}, Int. J. Mod. Phys. A \textbf{18}, 449 (2003); \\
{\tt http://gwdac.phys.gwu.edu/analysis/pin\_analysis.html}.
\bibitem{c:INFL}
I.G.~Alekseev \textit{et al.}, Phys. Rev. C \textbf{55}, 2049 (1997).
\bibitem{c:SETUP} 
Yu.A.~Beloglazov \textit{et al.}, Instrum. Exp. Tech. \textbf{47}, 744 (2005).
\bibitem{c:TARG}
E.I.~Bunyatova \textit{et al.}, \textit{Preprint LNPI-1191}, (LNPI, Leningrad 1986).
\bibitem{c:AEROGEL}
Yu.K.~Akimov \textit{et. al.}, Instrum. Exp. Tech. \textbf{45}, 634 (2005).
\bibitem{c:ITEP}
I.G.~Alekseev \textit{et al.}, Nucl. Phys. B \textbf{348}, 257 (1991).
\bibitem{c:ALBROWM}
M.G.~Albrow \textit{et al.}, Nucl. Phys. B \textbf{37}, 594 (1972).
\bibitem{c:YUKOSAWA}
D.~Hill \textit{et al.}, Phys. Rev. Lett. \textbf{27}, 1241 (1971).
\bibitem{c:MARTIN}
J.E.~Martin \textit{et al.}, Nucl. Phys. B \textbf{89}, 253 (1975).
\bibitem{c:ALBROWP}
M.G.~Albrow \textit{et al.}, Nucl. Phys. B \textbf{25}, 9 (1970).
\bibitem{c:EANDI}
R.D.~Eandi \textit{et al.}, Phys. Rev. \textbf{136}, B536 (1964).
\bibitem{c:BURLESON}
G.~Burleson \textit{et al.}, Phys. Rev. Lett. \textbf{26}, 338 (1971).


\end{thebibliography}
\end{document}